\documentclass{osa-article}

\journal{oe}
\articletype{Research Article}
\usepackage{dcolumn}
\usepackage{color}
\usepackage{bm}
\usepackage{siunitx}

\def\U#1{{\rm #1}} 
\def\u#1{_{\rm #1}}
\newcommand{\Arg}{\U{Arg}}
\renewcommand{\Re}{\operatorname{Re}}
\renewcommand{\Im}{\operatorname{Im}}

\begin{document}

\title{
Frequency stabilization via interference between transmitted and reflected lights from a reference cavity
}

\author{Rikizo Ikuta,\authormark{1,2,*}}

\address{\authormark{1}
Graduate School of Engineering Science, Osaka University, Toyonaka, Osaka 560-8531, Japan\\
\authormark{2}Center for Quantum Information and Quantum Biology, Osaka University, Toyonaka, Osaka 560-8531, Japan
}

\email{\authormark{*}ikuta@mp.es.osaka-u.ac.jp} 



\begin{abstract}
We propose a modulation-free optical frequency stabilization technique using an interferometric effect 
between transmitted and reflected lights from a reference cavity. 
The property of the reflected light brings robustness of the error signal against laser intensity fluctuations as in previous stabilization methods.
Due to the property of the transmitted light, 
the capture range for a specific locking frequency is expanded up to twice the FSR of the cavity, which we experimentally demonstrate. 
If locking to any resonant frequency is allowed, the capture range is infinite. 
From the effect of using both lights,
our method achieves the highest sensitivity to the frequency fluctuations around the resonant frequency 
and provides robustness against the interferometer fluctuations. 
\end{abstract}

%%%%%%%%%%%%%%%%%%%%%%%%%%  body  %%%%%%%%%%%%%%%%%%%%%%%%%%
\section{Introduction}
Stabilization of laser frequency is vital for various kinds of applications such as 
atomic clocks\cite{Takamoto2005,Jiang2011,Newman2019}, spectroscopy~\cite{Udem2002-1,Imai2016,Grinin2020},
and optical sensing~\cite{Dell2014, Shu2015, Asano2022}
as well as experiments for fundamental sciences~\cite{Pyka2013,Kumar2018,Ikuta2022-2}. 
Also recently, it is highly demanded in quantum information experiments 
aiming for quantum computers~\cite{Bruzewicz2019,Pogorelov2021,Bluvstein2022}, quanum sensing~\cite{Degen2017}, 
quantum key distribution~\cite{Lucamarini2018,Chen2022}, quantum internet~\cite{Maring2017,Ikuta2018,Liu2024} and so on. 
So far, 
significant efforts for developing techniques of optical frequency stabilization to a reference cavity have been made. 
The Pound-Drever-Hall~(PDH) method~\cite{Drever1983,Black2001} is one of the most prominent stablization methods. 
In PDH method, laser frequency is modulated to generate sidebands around the carrier frequency,
and then the modulated light is input to a reference cavity.
After the light reflected from the cavity is detected by a photodetector, 
the electrical signal is downconverted by mixing the local oscillator used for the modulation. 
Due to the high-frequency modulation technique, 
the downconverted signal used as the error signal of PDH method 
has limited low frequency errors such as DC offset and baseband noises.
In addition to the low noise properties,
the error signal has a feature of a sharp and antisymmetric shape around the resonant frequency 
with a capture range larger than the cavity linewidth.
While PDH method offers the high performances, 
there is no effective modulator system including the RF modulator and associated electronics 
working at all wavelengths with the desired modulation frequency, 
which implies there are cases where the PDH method may not be appropriate.
For example, frequency modulation in tens of GHz or higher will increase the complexity of electronics. 
The limit of the modulation frequency restricts the possible locking range~\cite{Zeyen2023}. 
An on-chip PDH system in the telecom band may be difficult due to the lack of efficient modulators~\cite{Idjadi2024}. 

As an alternative to PDH method, 
modulation-free stabilization methods which do not receive the restriction of the modulation systems 
have been actively studied~\cite{Hansch1980,Wieman1982,Shaddock1999,Sukenik2002,Diorico2024,Idjadi2024}. 
The Hansch-Couillaud~(HC) method~\cite{Hansch1980} 
is a representive method that has been commonly used~\cite{Scholz2007,Zhou2016,Li2018,Tinsley2021}. 
HC method uses a dispersive medium for separating resonant frequencies of horizontal~(H) and vertical~(V) polarized lights. 
By detecting an interference between the polarizing lights reflected by the cavity, 
the phase of the H-polarized light near the resonant frequency 
is extracted as a relative phase to that of the V-polarized light far from the resonant frequency. 
As a result, an error signal comparable to PDH method is obtained. 
The working mechanism of HC method was applied to methods using other degrees of freedom instead of polarization 
such as two optical paths~\cite{Wieman1982,Idjadi2024} and spatial TEM modes~\cite{Shaddock1999}. 
For example, in Ref.~\cite{Idjadi2024}, 
an interferometric field between light coupled to a microring cavity and uncoupled light is detected 
at each output port of the Mach-Zehnder interferometer~(MZI). 
Difference of the detected signals yields an error signal comparable to PDH method 
without any modulation and dispersive medium.
Hereafter, we collectively refer to methods equivalent to the above interferometric method as MZI methods. 
While MZI methods are sensitive to the low-frequency noise and DC offsets, different from PDH method, 
usefulness of MZI methods with reproducing the performance of PDH method has been experimentally shown.
However, apart from advantages of the implementation, 
functional novelties enabled by eliminating RF modulation have not been demonstrated. 

In this paper, we propose a modulation-free laser frequency stabilization method 
that essentially shows a performance improvement compared to the previous methods. 
The distinctive feature of our method is the utilization of both reflected and transmitted light from a reference cavity. 
The method using interference between them 
achieves the theoretical limit of sensitivity of the error signal around the resonant frequencies. 
The property of transmitted light expands the capture range for a locking frequency up to twice the FSR, 
surpassing the FSR achieved by previous methods based on only reflected light,
which we experimentally demonstrate.
If the locking point is not restricted to a specific frequency, the capture range is infinite 
without limitation of RF modulation~\cite{Zeyen2023}. 
In addition to the above features, 
the property of reflected light brings the robustness of the error signal against laser intensity fluctuations 
comparable to the previous methods. 
Our method is insensitive to the interferometer fluctuations and thus stably hold the above properties of the error signal. 

\section{Theory}
\subsection{An error signal based on transmitted and reflected lights from a cavity}
\begin{figure}[t]
 \begin{center}
   \scalebox{1}{\includegraphics[bb=0 0 196 111]{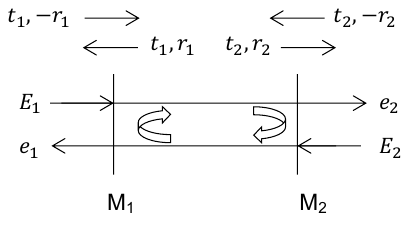}}
      \caption{
        The Fabry-P\'{e}rot~(FP) cavity.
        $|t_{1(2)}|^2$ and $|r_{1(2)}|^2$ are transmittance and reflectance of mirror $\U{M}\u{1(2)}$, respectively. 
      }
 \label{fig:cavity}
 \end{center}
\end{figure}
We explain the theory of our stabilization method by analyzing lights transmitted and reflected from a reference cavity.
We consider the Fabry-P\'{e}rot~(FP) cavity here,
but other cavities will be used if both transmitted and reflected lights can be obtained 
such as a microring cavity coupled to two optical fibers. 
The FP cavity is composed of two mirrors $\U{M}_{1}$ and $\U{M}_{2}$ as shown in Fig.~\ref{fig:cavity}. 
Supposing that vectors of input and output optical fields labelled $1$ and $2$ are 
$(E_1,E_2)$ and $(e_1,e_2)$, respectively, the transformation matrix $U$ of the FP cavity
from the input to the output fields is written by 
\begin{align}
U = 
  \begin{pmatrix}
    u_{11} & u_{12} \\
    u_{21} & u_{22} 
  \end{pmatrix},
  \label{eq:unitary}
\end{align}
where $u_{OI}$ for $I=1,2$ and $O=1,2$ is the transition amplitude from input port $I$ to output port $O$.
Using amplitudes of reflectance and transmittance of the mirrors depicted in Fig.~\ref{fig:cavity}, 
in which the four amplitudes are assumed to be real values for simplicity, 
the matrix elements are described by 
\begin{align}
u_{11} &= -r_{1}+\frac{t_{1}^2r_{2}e^{i\theta}}{1-r_{1}r_{2}e^{i\theta}},\qquad
u_{22} &= -r_{2}+\frac{t_{2}^2r_{1}e^{i\theta}}{1-r_{1}r_{2}e^{i\theta}},\qquad
u_{12} &= u_{21} = \frac{t_{1}t_{2}e^{i\theta/2}}{1-r_{1}r_{2}e^{i\theta}}. 
\label{eq:ulr}
\end{align}
$\theta = 2nL\omega/c$ is a round-trip phase shift 
determined by cavity length $L$, speed of light $c$, refractive index $n$, and angular frequency of light $\omega$. 

When we input an optical light $(E_1,E_2)=(1,e^{i\phi})/\sqrt{2}$ to the cavity
with omitting the phase rotating at $\omega$, 
we obtain $e_1$ and $e_2$ coming from the left and right ends of the cavity as
\begin{align}
  \begin{pmatrix}
    e_1\\
    e_2
  \end{pmatrix}
  = U
    \begin{pmatrix}
    E_1\\
    E_2
    \end{pmatrix}
  =\frac{1}{\sqrt{2}}
  \begin{pmatrix}
  u_{11} + e^{i\phi} u_{12} \\
  u_{21} + e^{i\phi} u_{22} 
  \end{pmatrix}. 
  \label{eq:e1e2}
\end{align}
\begin{figure}[t]
 \begin{center}
   \scalebox{1}{\includegraphics[bb=0 0 381 140]{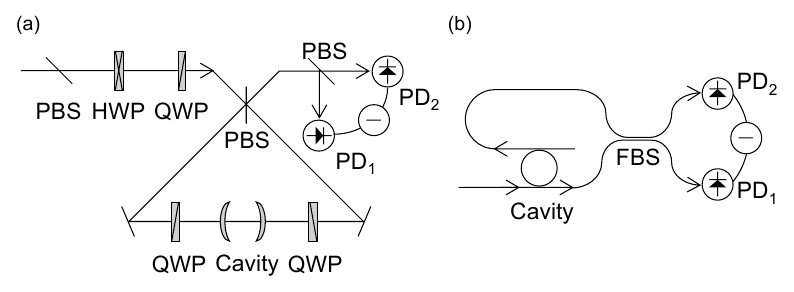}}
    \caption{
        Experimental setups.
        (a)
      The signal light is input to the reference cavity from CW and CCW directions. 
      Two QWPs inside the interferometer are $\pi/4$ rotated with their rotating directions opposite to each other. 
      (b)
      Implementation using a ring cavity. A fiber-based beamsplitter~(FBS) mixes the
      the transmitted and reflected lights from the cavity with the relative phase of $\phi$.
      }
 \label{fig:setup}
 \end{center}
\end{figure}
The optical fields $e_1$ and $e_2$ can be extracted deterministically, 
which we show by concretely constracting an optical circuit depicted in Fig.~\ref{fig:setup}~(a). 
We prepare a polarizing input light described by the Jones vector 
$(E_1,E_2)=(1,e^{i\phi})/\sqrt{2}$ with relative phase $\phi$ between H and V
using a pair of a half-wave plate~(HWP) and a quarter-wave plate~(QWP). 
Two QWPs inserted in the interferometer are $\pi/4$ rotated 
with their rotating directions opposite to each other. 
H- and V-polarized components of the input light are separated by a polarizing beamsplitter~(PBS)
and coupled to the interferometer from clockwise~(CW) and 
counter-clockwise~(CCW) directions, respectively. 
For CW~(CCW) direction,
transmitted and reflected lights from the cavity are extracted out of the interferometer as H(V)- and V(H)-polarized lights. 
Consequently, lights coming from two output ports of second PBS after the interferometer correspond to $e_1$ and $e_2$. 
Using two photodetectors $\U{PD}_1$ and $\U{PD}_2$, we obtain difference of the intensities as 
\begin{align}
  S&=|e_2|^2-|e_1|^2\\
   &=  \frac{1}{2}\left(
     |u_{21}(\theta) + e^{i\phi} u_{22}(\theta)|^2 - |u_{11}(\theta) + e^{i\phi} u_{12}(\theta)|^2
     \right)\\
  &= -2\Re[e^{i\phi} u_{11}^*(\theta) u_{12}(\theta)]. 
  \label{eq:S}
\end{align}
Here we used $u_{12}=u_{21}$ and
properties of unitary matrix $U$ as $|u_{11}|^2+|u_{12}|^2=|u_{21}|^2+|u_{22}|^2$
and $u_{11}^*u_{12}+u_{21}^*u_{22}=0$. 

Another setup for obtaining $S$ is realized 
by moving the HWP and QWP before the interferometer to the end just before the third PBS, 
using an input signal of $(E_1,E_2)=(1,0)$.
In the case, the optical field after the interferometer becomes $(u_{21}, u_{11})$.
The error signal obtained after mixing the polarization using the waveplates with appropriate settings 
followed by the third PBS is $-2\Re[e^{i\phi} u_{11}^*(\theta) u_{21}(\theta)]$,
which is exactly the same as $S$ in Eq.~(\ref{eq:S}) because of $u_{21}=u_{12}$. 
A similar setup could be realized by using a ring cavity coupled to two optical fibers 
and the MZI based on two spatial modes, as shown in Fig.~\ref{fig:setup}~(b).
While the interferometer using both transmitted and reflected lights may be larger than that using only the reflected light, the on-chip integration using a ring cavity will be useful to operate the circuit stably like Ref.~\cite{Idjadi2024}. 
In addition, this circuit without the polarization degree of freedom would work well without suffering from the birefringence of the cavity. 

\subsection{Properties of the error signal using a symmetric cavity, as the ideal case}
When the cavity is symmetric as $r=r_1=r_2$ and $t=t_1=t_2$,
$u_{11}^*u_{12}$ becomes the pure imaginary. Then $S$ is written by 
\begin{align}
  S &=-2\sin\phi |u_{11}(\theta) u_{12}(\theta)|\sin(\Arg (u_{11}(\theta)) - \Arg (u_{12}(\theta))) \label{eq:Sarg}\\
   &=-4\cos\Delta \frac{t^2r\sin(\theta/2)}{1-2r^2\cos\theta+r^4}, 
     \label{eq:Ssym}
\end{align}
where $\Delta = \phi - \pi/2$ is a deviation from an optimal value $\phi=\pi/2$ for frequency stabilization. 

\begin{figure}[t]
 \begin{center}
   \scalebox{1}{\includegraphics[bb=0 0 215 147]{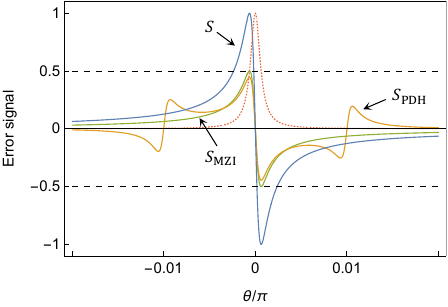}}
      \caption{
        Error signals $S$~(blue), 
        $S\u{PDH}$~(orange), 
        and $S\u{MZI}$~(green) 
        with mirror parameters of $r_1=r_2=0.999$ and $t_1=t_2=\sqrt{1-r_1^2}$.
        For PDH method, $\beta=1/2$ and $\theta\u{P}=\pi/10^2$ is used. 
        $\phi=\pi/2~(\Delta=0)$ is chosen to maximize the error signals for the other two methods. 
        A red dashed curve is transmission spectrum of the cavity. 
      }
 \label{fig:error}
 \end{center}
\end{figure}
For the ideal case of $\Delta=0$, error signal $S$ is described in Fig.~\ref{fig:error}. 
As references, we show error signals obtained by PDH and MZI methods using the symmetric cavity,
and transmission spectrum $|u_{12}|^2$ of the cavity. 
The error signals used for the simulation are summarized as follows~\cite{Black2001,Idjadi2024}:
\begin{align}
  S\u{PDH} &= \frac{2\beta}{2+\beta^2} \Im [u_{11}(\theta)u_{11}^*(\theta+\theta\u{P})
-u_{11}^*(\theta)u_{11}(\theta-\theta\u{P})],
             \label{eq:PDH}\\
  S\u{MZI} &= \frac{1}{4}(|u_{11}(\theta) - e^{i\phi}|^2 - |u_{11}(\theta) + e^{i\phi}|^2)\\
           &= -\Re[e^{i\phi}u_{11}^*(\theta)]\\
           &= - |u_{11}(\theta)| \sin(\Arg (u_{11}(\theta))-\Delta). 
             \label{eq:MZI}
\end{align}
$\beta\ll 1$ is the modulation index of an electro optic modulator.
$\theta\u{P}$ is the phase shift corresponding to the sideband frequency. 
We treated the transmitted light coupled to the microring cavity as the reflected light by the FP cavity. 
The error signal in HC method is 
\begin{align}
  S\u{HC}
  &= \frac{1}{4}(|u_{11}(\theta) + e^{i\phi}u_{11}(\theta+\theta_{H})|^2
    - |u_{11}(\theta) - e^{i\phi}u_{11}(\theta+\theta_{H})|^2)\\
  &= \Re[e^{i\phi}u_{11}^*(\theta)u_{11}(\theta+\theta\u{H})]\\
  &\sim S\u{MZI}, 
    \label{eq:HC}
\end{align}
where $\theta\u{H}$ is the phase difference between H and V due to dispersion. 
The approximation to $S\u{MZI}$ is obtained by using $u_{11}(\theta+\theta_{H}) \sim -r_{1}$ around the resonant frequency 
as described in Ref.~\cite{Hansch1980} and $r_{1}\sim 1$. 
\begin{figure}[t]
 \begin{center}
   \scalebox{1}{\includegraphics[bb=0 0 216 218]{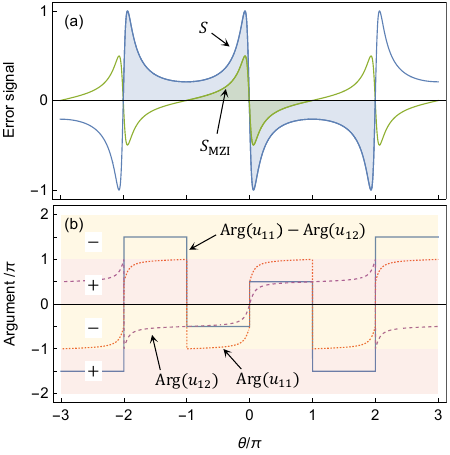}}
      \caption{
        (a)
        Error signals $S$ and $S\u{MZI}$ with $r_1=r_2=0.9$, $t_1=t_2=\sqrt{1-r_1^2}$ and $\phi=\pi/2$.
        Filling regions show capture ranges of the lock to the frequency at $\theta=0$.
        (b)
        Arguments of transmitted and reflected lights, and their difference. 
        Plus and minus regions describe the signs of the sine functions 
        which determine signs of the error signals. 
      }
 \label{fig:error2}
 \end{center}
\end{figure}
Indeed the detailed behivior of $S\u{HC}$ is different from $S\u{MZI}$ for frequencies far from the resonant frequency 
or for $\Delta\neq 0$, 
but the difference is not so significant in terms of comparison of these methods to our method discussed below. 

From Fig.~\ref{fig:error}, all methods show sharp and antisymmetric error signals around the resonant frequency. 
The error signals take zero at $\theta=0$,
which means all methods are robust against intensity fluctuations of the input light. 
For the sensitivity, $S$ is the highest achieving theoretical limit 
from $-1$ to $1$ around the resonant frequency. 
The height of $S\u{PDH}$ is unavoidably reduced because of the modulation determined by $\beta$. 
For $S\u{MZI}$, because half of the total intensity escapes as the transmitted light, the sensitivity is dropped to half. 
While a larger value of $r_2^2(>r_1^2)$ gives a larger sensitivity, 
an extremely asymmetric cavity would not be conventionally used 
because the transmitted light is used for monitoring the mode matching to the cavity. 
In contrast, all components of the light are used for the error signal in our method, and the highest sensitivity is achieved. 

Next, to see the wide capture range of $S$, we depict $S$ and $S\u{MZI}$ in Fig.~\ref{fig:error2}~(a). 
$S\u{MZI}$ is repeated at the cycle of FSR because the method uses only reflected light. 
From Eq.~(\ref{eq:MZI}), the sign of $S\u{MZI}$ is determined by that of $\sin(\Arg (u_{11}))$ for $\Delta=0$ 
which is reversed every $\pi$ period of $\theta$ as shown in Fig.~\ref{fig:error2}~(b). 
This limits their capture range to be smaller than FSR. It is the same for PDH, MZI and other methods 
based on only the reflected light. 
However, for $S$, the repeated cycle is twice the FSR corresponding to $4\pi$ rotation of $\theta$. 
This is because our method uses the transmitted light 
which includes the component of half phase change of the reflected light as in Eq.~(\ref{eq:Ssym}). 
The visual understanding of the sign of $S$ characterized by $\sin(\Arg (u_{11}) - \Arg (u_{12}))$ 
is in Fig.~\ref{fig:error2}~(b). 
We see the sign of the function changes for each $2\pi$ rotation of $\theta$. 
As a result, the capture range of $S$ to the frequency at $\theta=0$ in Fig.~\ref{fig:error2}~(a) is achieved to be twice the FSR.
The minimal value of $|S|$ is $\sim 2(1-|r|)$ at $\theta=\pm \pi$. 
If any resonant frequencies are allowed to use as locking points, 
$S$ has an infinite capture range of the lock. 

\begin{figure}[t]
 \begin{center}
   \scalebox{1}{\includegraphics[bb=0 0 216 147]{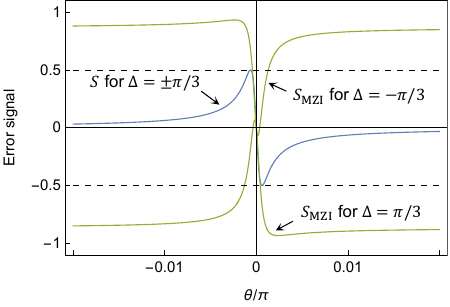}}
      \caption{
        Error signals $S$ and $S\u{MZI}$ for $\Delta=\pm \pi/3$. 
        Other parameters are the same as those used in Fig.~\ref{fig:error}. 
      }
 \label{fig:phase}
 \end{center}
\end{figure}
Finally, we show robustness of $S$ against fluctuation of $\phi(=\Delta +\pi/2)$. 
Different from the original Sagnac interferometer, 
the optical paths of the transmitted and reflected lights in Fig.~\ref{fig:setup} are not completely shared, 
which may cause nonzero value of $\Delta$ due to the path length fluctuations. 
When we explicitely denote the $\Delta$ dependency of $S$ as $S=S(\Delta)$,
Eq.~(\ref{eq:Ssym}) is written by $S(\Delta)=\cos\Delta\cdot S(0)$, in which 
the effect of the fluctuations appears in the coefficient of $\cos\Delta$. 
Thus, the shape of $S$ including the zero offset at $\theta=0$
and the wide capture range hold regardless of the fluctuations,
even though it does not use copropagating light. 
For $|\Delta| < \pi/3$, more than half the maximum height remains, which we show in Fig.~\ref{fig:phase}. 
To stabilize the interferometer for maximizing the height of the error signal,
an active feedback to the mirror inside the interferometer
using $|e_{1(2)}|^2$ with electrically removing the offset or error signal $S$ might be used. 

The reason for the high robustness of $S$ is 
separating the contribution of $\Delta$ from $\Arg (u_{11})-\Arg (u_{12})$. 
For $S\u{MZI}$, because contributions of $\Delta$ and $\Arg (u_{11})$ are not separated as in Eq.~(\ref{eq:MZI}), 
nonzero values of $\Delta$ shift the value of $\theta$ where the sign of the error signal is reversed. 
This effect makes the shape of $S\u{MZI}$ less antisymmetric which we show in Fig.~\ref{fig:phase}. 
In our method, by using both the reflected and transmitted lights, 
the unitary nature of the cavity works effectively, 
resulting in the successful separation of contributions of $\phi$
and pure imaginary $u_{11}^*u_{12}$ in Eq.~(\ref{eq:S}). 

\subsection{Effect of the asymmetry of the reflectances on the error signal}
\begin{figure}[t]
 \begin{center}
   \scalebox{1}{\includegraphics[bb=0 0 365 134]{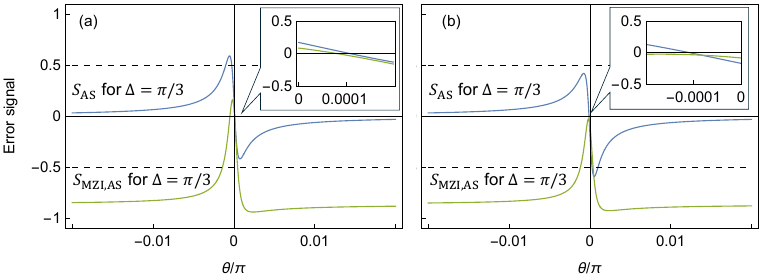}}
      \caption{
        Error signals $S\u{AS}$ and $S\u{MZI,AS}$ for $\Delta=\pi/3$ with
        (a) $r_1=0.9989$ and $r_2=0.9991$ and
        (b) $r_1=0.9991$ and $r_2=0.9989$.
        For $\Delta=-\pi/3$, the point-symmetric curves with respect to the origin are obtained.
      }
 \label{fig:drift}
 \end{center}
\end{figure}
  In the previous subsection, we have discussed the properties of the locking system using the symmetric cavity 
  which is the main focus of this paper. 
It is important to note that the robustness against $\Delta$ in our method is based on the symmetry of the cavity.
For $r_1\neq r_2$, the property does not hold, 
whereas the other properties of the high sensitivity and the wide capture range, 
brought by the unitary transformation of the cavity, remain. 
In the following,
we see the effect of the difference between $r_1$ and $r_2$ on the error signal in more detail. 
We suppose $r_2=(1+\epsilon)r_1$ for $|\epsilon| \ll 1$. 
In this case, the error signal is described by 
\begin{align}
  S\u{AS}
  &= \cos\Delta \cdot \tilde{S}
    - \sin\Delta \frac{2(1-r^2)r\cos(\theta /2)\epsilon}{1-2r^2\cos\theta +r^4} +\mathcal{O}(\epsilon^2),
\end{align}
where $r=r_1$ and 
\begin{align}
  \tilde{S}&=S(0)
             \left(
             1 + \frac{(1+r^2)(1-4r^2+2r^2\cos\theta +r^4)\epsilon}{2(1-r^2)(1-2r^2\cos\theta+r^4)}
             \right). 
\end{align}
Here $S(0)$ is the error signal obtained when $r_1=r_2=r$ and $\Delta=0$ are satisfied in Eq.~(\ref{eq:Ssym}).
The dominant portion of the error signal $S\u{AS}$ is $S(0)$, 
which means the property of the asymmetric signal around the locking point with the high sensitivity 
remains. 
For the robustness against the path length fluctuations, due to the component related to $\sin\Delta$, 
$\Delta\neq 0$ causes the drift of the locking point. 
Thus, a careful stabilization of the interferometer like other MZI methods will be required. 

Examples of the error signals at $\Delta=\pi/3$
for $\epsilon=\pm 0.0002$ with the cavity finesse corresponding to the case in Fig.~\ref{fig:error2}
are shown in Figs.~\ref{fig:drift}~(a) and (b).
As a reference, we plot the error signal $S\u{MZI,AS}$ obtained by MZI method with the asymmetric mirrors.
Apart from the drift described above, 
we see that the behavior of the two methods is very different, 
similar to the result in Fig.~\ref{fig:phase}.
The error signal $S\u{MZI,AS}$ based on the light coming from only one side of the cavity 
has the different shapes for $r_1 < r_2$ and $r_1 > r_2$.
As seen in Fig.~\ref{fig:drift}~(b),
a serious situation where $S\u{MZI,AS}$ does not reach zero may occur,
which means the frequency locking cannot be achieved.  
In contrast, our method has the same shape of the error signals for the two cases 
while keeping the nearly antisymmetric property with respect to the locking point.
In particular, the wide capture range reaching twice the FSR holds not only in this case but in general. 
To see this, we describe the error signal $S\u{AS}$ in the asymmetric reflectances
using Eq.~(\ref{eq:S}) as 
\begin{align}
  S\u{AS} &= -2|u_{11}(\theta) u_{12}(\theta)|\sin(\Arg (u_{11}(\theta)) - \Arg (u_{12}(\theta))-\Delta). 
\label{eq:Sas}
\end{align}
\begin{figure}[t]
 \begin{center}
   \scalebox{1}{\includegraphics[bb=0 0 371 214]{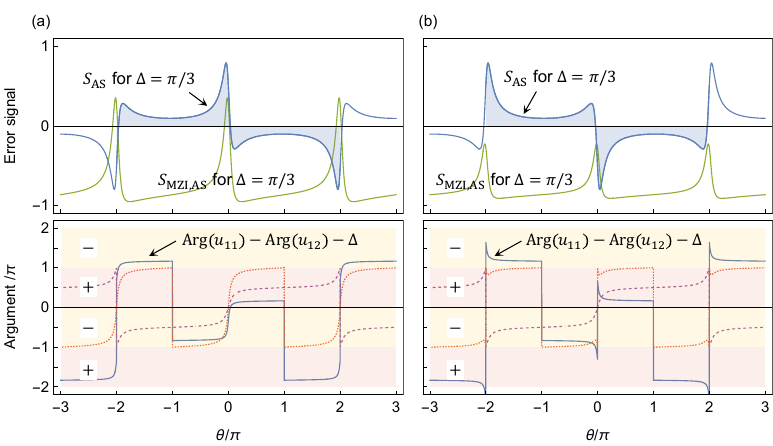}}
      \caption{
        Error signals for $\Delta=\pi/3$ and the related arguments 
        with 
        (a) $r_1=0.87$ and $r_2=0.93$ and (b) $r_1=0.93$ and $r_2=0.87$.
        Dotted curves show $\Arg(u_{12})$~(purple) and $\Arg(u_{11})$~(orange).
      }
 \label{fig:cap2}
 \end{center}
\end{figure}
From Eq.~(\ref{eq:Sas}), the capture range is determined by the values of $\theta$
where the sine function crosses zero.
For visual understanding,
we use Figs.~\ref{fig:cap2}~(a) and (b) which show examples of the capture range and the related arguments, 
using parameters giving the finesse corresponding to that in Fig.~\ref{fig:error2}. 
As seen in lower figures in Figs.~\ref{fig:cap2}~(a) and (b),
while the shape of $\Arg (u_{11}) - \Arg (u_{12})$ depends on the relationship between $r_1$ and $r_2$, 
it keeps the minus~(plus) sign of the sine function for $-2\pi < \theta \leq 0~(0 \leq \theta < 2\pi)$. 
In addition, from the phase property of the reflected light characterized by $u_{11}$ in Eq.~(\ref{eq:ulr}), 
we see the value of $\Arg (u_{11}) - \Arg (u_{12})$ shifts discontinuously at $\theta =\pm \pi$, 
while keeping the sign of the sine function. 
A similar discontinuous shift appears at $\theta=0$ for $r_1 > r_2$ as in Fig.~\ref{fig:cap2}~(b). 
On the other hand, the value of $\Arg (u_{11}) - \Arg (u_{12})$ around $\theta=0$ changes continuously for $r_2 > r_1$ as in Fig.~\ref{fig:cap2}~(a). 
Thus, in any cases, the path length fluctuations for $|\Delta|\leq \pi/2$ do not 
increase the number of solutions for $\theta$ where the sine function equals zero,
although there is a change in the value of the solution itself, which is observed as a drift of the locking point.
Consequently, the error signal remains a wide capture range property,
unlike MZI methods that are sensitive to path length fluctuations, even if the reflectances are unbalanced.
The property is satisfied generally, which is confirmed by direct calculation using Eq.~(\ref{eq:ulr}). 
This fact will allow the interferometer to be stabilized while maintaining the frequency lock of the target light. 

\section{Experiment }
\begin{figure}[t]
 \begin{center}
   \scalebox{1}{\includegraphics[bb=0 0 221 220]{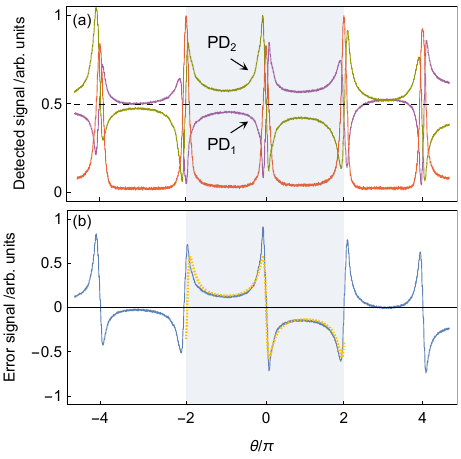}}
    \caption{
      (a) Observed signals by $\U{PD}_1$ and $\U{PD}_2$ corresponding to $|e_1|^2$ and $|e_2|^2$.
      They are normalized by the peak value in the upper left at $\theta=0$ of the signal by $\U{PD}_2$. 
      Red data is the transmitted spectrum detected by $\U{PD}_1$ for the input of H-polarized light to the interferometer. 
      (b) Experimentally obtained error signal~(blue) by taking difference between signals 
      detected by $\U{PD}_2$ and $\U{PD}_1$ in Fig.~\ref{fig:result}~(a). 
      A dashed curve~(yellow) is a fitting function proportional to Eq.~(\ref{eq:Ssym}). 
    }
 \label{fig:result}
 \end{center}
\end{figure}

For the proof of concept of our method,
we constract an experimental setup based on Fig.~\ref{fig:setup}.
Using the error signal obtained by the setup, we show the wide capture range over FSR in our method which never been observed in other methods based on the reflected light. 
The wavelength of input light is \SI{1540}{nm}. 
As a reference cavity, we used a solid etalon with the length of \SI{13}{mm} corresponding to FSR of \SI{8}{GHz}.
To select the single longitudinal mode of the cavity, the lights after second PBS are coupled to single-mode fibers followed by PDs. 
From the transmitted spectra, one of which is shown in Fig.~\ref{fig:result}~(a), 
the finesse of the cavity is estimated to be 13.0, which corresponds to the refrectance of $r^2=0.79~(r=0.89)$. 
By setting phase $\phi$ to maximize the sensitivity of the error signal, 
$|e_1|^2$ and $|e_2|^2$ are experimentally obtained by $\U{PD}_1$ and $\U{PD}_2$, respectively, 
as shown in Fig.~\ref{fig:result}~(a).
Taking their difference, we obtain the error signal $S$ depicted in Fig.~\ref{fig:result}~(b). 
Within the range of $-2\pi\leq \theta \leq 2\pi$, 
we clearly see that the capture range of the error signal related to the resonant frequency at $\theta=0$ is close to twice the FSR. 
For this range, the best fit to the data with a function proportional to Eq.~(\ref{eq:Ssym}) gives $r=0.89$, 
which is in good agreement with the observed value.
We thus conclude the error signal based on our method is successfully demonstrated. 
The error signal for $|\theta| > 2\pi$ does not keep the desired shape observed within $|\theta| \leq 2\pi$. 
We guess the reason may be the dispersion effect of the solid etalon and other optical components 
which would cause the frequency dependency of $\phi$. 

\section{Conclusion}
In conclusion, we proposed a modulation-free optical frequency stabilization method 
based on transmitted and reflected lights from a reference cavity. 
The error signal using them inherits both properties of error signals based on only one of them. 
Namely, the error signal in our method is not affected by laser intensity fluctuations 
and is allowed to have the capture range beyond FSR of the cavity. 
The latter property leads to an infinite capture range of the error signal for locking on the resonant frequencies.
As a property arising from an interferometric effect between the transmitted and reflected light, 
the error signal has the largest sensitivity to frequency fluctuations around the resonant frequency. 
In addition, the robustness against the interferometer fluctuations is achieved. 
We experimentally demonstrated the acquisition of the error signal based on the proposed method,
and showed the capture range reaching twice the FSR. 
The stabilization method will be helpful in obtaining narrow linewidth lasers as a complementary method to conventional methods,
especially when using a relatively low-power signal light and/or a low-finesse and small cavity. 

The method uses the phase difference between lights coming from two output ports of the cavity in Fig.~\ref{fig:cavity}. 
Other cavity types such as confocal cavities, 
triangle cavities and so on could be used to obtain lights with phase relations different from that in this paper. 
Useful error signals or other applications may be found by interferences between them. 
In the experiment, the input light is coupled to both ends of the cavity and both transmitted and reflected lights are used. 
We believe such usage of the cavity as a frequency sensitive beamsplitter unlike conventional broadband beamsplitters 
will lead to emergence of new optical technologies. 

\section*{Funding}
This work was supported by Moonshot R \& D, JST JPMJMS2066 and JST JPMJMS226C; 
FOREST Program, JST JPMJFR222V; 
R \& D of ICT Priority Technology Project JPMI00316, and Asahi Glass Foundation. 

\section*{Acknowledgments}
We thank Masao Kitano for many fruitful discussions about the research 
and thank Hiroki Takahashi and Yoshiaki Tsujimoto for valuable comments on the paper draft. 

\section*{Disclosures}
The authors declare no conflicts of interest.

\section*{Data availability}
Data available from the authors on request.

\end{document}